\documentclass[aps,prd,preprint,superscriptaddress,showpacs]{revtex4}
\usepackage{amsmath}
\usepackage{amssymb}
\usepackage{epsfig}
\usepackage{ulem}
\usepackage{graphics}
\usepackage{indentfirst}
\usepackage{color}

\newcommand{\tb}{\ensuremath{\bar{T}}}
\newcommand{\kb}{\ensuremath{\bar{\chi}}}
\newcommand{\pb}{\ensuremath{\bar{\phi}}}

\newcommand{\lml}{\ensuremath{\lambda_1}}
\newcommand{\lmll}{\ensuremath{\lambda_2}}

\newcounter{RomanNumber}

\begin{document}

\preprint{ACT-7-14, MIFPA-14-16}

\title{Natural Inflation with Natural Trans-Planckian Axion Decay Constant from Anomalous $U(1)_X$}

\author{Tianjun Li \vspace*{0.5cm}}
\affiliation{{\small State Key Laboratory of Theoretical Physics
and Kavli Institute for Theoretical Physics China (KITPC),
      Institute of Theoretical Physics, Chinese Academy of Sciences,
Beijing 100190, P. R. China}}

\affiliation{{\small School of Physical Electronics,
University of Electronic Science and Technology of China,
Chengdu 610054, P. R. China}}

\author{Zhijin Li}

\affiliation{{\small George P. and Cynthia W. Mitchell Institute for
Fundamental Physics and Astronomy,
Texas A\&M University, College Station, TX 77843, USA}}

\author{Dimitri V. Nanopoulos}

\affiliation{{\small George P. and Cynthia W. Mitchell Institute for
Fundamental Physics and Astronomy,
Texas A\&M University, College Station, TX 77843, USA}}

\affiliation{{\small Astroparticle Physics Group, Houston Advanced
Research Center (HARC), Mitchell Campus, Woodlands, TX 77381, USA}}

\affiliation{{\small Academy of Athens, Division of Natural Sciences,
28 Panepistimiou Avenue, Athens 10679, Greece} \vspace*{1.6cm}}

\begin{abstract}

We propose a natural inflation model driven by an imaginary or axionic component
of a K\"ahler modulus in string-inspired supergravity. The shift symmetry of the
axion is gauged under an anomalous $U(1)_X$ symmetry, which leads to a modulus-dependent
Fayet-Iliopoulos (FI) term. The matter fields are stabilized by F-terms, and
the real component of the modulus is stabilized by the $U(1)_X$ D-term, while
its axion remains light. Therefore, the masses of real and imaginary components of
the modulus are separated at different scales. The scalar potential for natural inflation
is realized by the superpotential from the non-perturbative effects. The trans-Planckian
axion decay constant, which is needed to fit with BICEP2 observations, can be
obtained naturally in this model.

\end{abstract}


%\pacs{11.25.-w, 98.80.-k}
\maketitle

\section{Introduction}

The recent discovery of the B-mode polarization by the BICEP2 Collaboration, if confirmed, provides
further strong evidence on the inflationary paradigm for the early epoch of the Universe \cite{Ade:2014xna}.
The observed tensor-to-scalar ratio $r$ is $r=0.20^{+0.07}_{-0.05}$, or $r=0.16^{+0.06}_{-0.05}$ without
the dust contributions. One of the inflation models, which agrees with the BICEP2 results, is
the well-known natural inflation \cite{Freese:1990rb}.

The motivation of natural inflation \cite{Freese:1990rb} is to solve the flatness problem of inflation potential
 at tree level, and remains flat against radiative corrections. The continuous shift symmetry protects the flatness
of inflation potential. To realize inflation, the continuous symmetry should be broken to a discrete shift symmetry
$\phi\rightarrow \phi+2\pi f$ with $f$ the axion (or inflaton) decay constant, and
the potential for natural inflation is
\begin{equation}
V(\phi)=\Lambda^4(1\pm{\rm cos}(\frac{\phi}{f}))~,~ \label{V}
\end{equation}
where $\Lambda$ is the inflation energy scale around $2\times 10^{16}$ GeV or
$10^{-2} ~{\rm M_{Pl}}$ for $r=0.16/0.20$ with ${\rm M_{Pl}}$ the reduced Planck mass
(${\rm M_{Pl}}=2.4\times 10^{18}$ GeV).

Axion is a ``natural" inflaton candidate since it preserves the exact continuous shift symmetry at perturbative level.
Axions can be obtained from antisymmetric tensor fields in string theory through spacetime
compactification \cite{Svrcek:2006yi}. Considering the non-perturbative effects, such as gaugino condensation
or instanton effect, one can break such continuous shift symmetry of axion to the discrete symmetry
 and then realize natural inflation in string theory.

The challenge to natural inflation is the large decay constant $f$. To generate sufficient large tensor fluctuations
that are consistent with BICEP2 observations, the decay constant $f$ should be trans-Planckian $f\sim O(10)$
in the Planck units \cite{Freese:2014nla}. However, in string theory the axion decay constant cannot be larger
than the string scale ${\rm M_{String}}$ \cite{Svrcek:2006yi, Choi:1985je, Banks:2003sx}, which is about one order
below the reduced Planck scale as required by the weak interaction assumption.
An effective large axion decay constant
was realized by the N-flation~\cite{Liddle:1998jc, Dimopoulos:2005ac} or
the aligned axion mechanism \cite{Kim:2004rp}. In the aligned natural inflation~\cite{Kim:2004rp},
two axions with small
decay constants are carefully adjusted to form a flat direction for inflation, and then one-linear combination
of two axions can have the effective trans-Planckian decay constant. Recently, there are
many works proposed to realize large decay constant with multi-aligned axions \cite{Choi:2014rja} or realize N-flation and
natural inflation in
string theory~\cite{Cicoli:2014sva, Marchesano:2014mla, Hebecker:2014eua, Blumenhagen:2014gta, Grimm:2014vva, Kallosh:2014vja, Arends:2014qca}.

To obtain inflation in string theory or its low energy approximation--supergravity (SUGRA), there is a general problem
on the moduli stabilization. For single field inflation, all the scalars except the inflaton should be fixed during
inflation. The well-known KKLT mechanism based on F-term was proposed in \cite{Kachru:2003aw}, where
 the complex-structure moduli are stabilized by the fluxes \cite{Giddings:2001yu} while the K\"ahler modulus
is stabilized through the non-perturbative effects. The difficulty to realize axion inflation based on
the KKLT mechanism is: once the real component of the K\"ahler modulus is fixed, its axionic component
obtains large mass as well and then destroys inflation \cite{Kallosh:2007ig}. This problem can be solved
by considering the modulus-dependent FI term associated with anomalous $U(1)_X$ \cite{Li:2014owa, Li:2014lpa}.
The FI term only depends on the real component of modulus, which can obtain large mass from D-term flatness.
The axion is still light as it decouples from D-term.

In this work, we construct a natural inflation model in string-inspired SUGRA where only one modulus couples
to the matter fields. The shift symmetry of axion is gauged to obtain anomalous $U(1)_X$. The gauge invariant
superpotential consists of the non-perturbative term of modulus and various couplings among matter fields.
All the matter fields are stabilized by F-terms, and the vanishing D-term gives large mass to the real component
of the modulus. The axion is still light after modulus stabilization and its potential is given exactly
by Eq.~(\ref{V}). Besides, the trans-Planckian axion decay constant can be obtained naturally
 by taking the large condensation gauge group and numbers of $U(1)_X$ charged matter fields (around 20).

This paper is organized as follows. In Section 2 we provide the string-inspired SUGRA structure for model building
and show the anomaly cancellation of $U(1)_X$. In Section 3 the matter fields/modulus stabilizations based on
the F-terms/D-term are discussed. In Section 4 we obtain the natural inflation potential after stabilizations,
and the trans-Planckian axion decay constant is realized from the non-perturbative effects
with suitable condensation gauge group. We conclude in Section 5.

\section{Natural Inflation Model Building}

We consider the following K\"ahler potential
\begin{equation}
K=-{\rm ln}(T+\tb)+\phi_i\pb_i+\chi_j\kb_j+\varphi_k\bar{\varphi}_k+Q\bar{Q}+X_l\bar{X}_l~,
\end{equation}
in which the indexes are $i=1,2, \cdots, m$, $j=1,2,\cdots,m-2$, $k=1,2,3,4$ and $l=0,1,2,\cdots,m$. The modulus $T$ can be dilaton superfield or one of the K\"ahler moduli from string compactification. The $U(1)_X$ charged matter fields
$(\phi_i, \chi_j, \varphi_k, Q)$, which will be generically denoted as $z_h$, are introduced
to construct gauge invariant superpotential and cancel the gauge anomalies.
The modulus $T$ and matter fields $z_h$ transform under $U(1)_X$ as follows
\begin{equation}
\begin{split}
T\rightarrow T+i\delta \epsilon ~\,\\
z_h\rightarrow z_h e^{i\epsilon q_{z_h}}~,
\end{split}
\end{equation}
in which the $U(1)_X$ charges $q_z$ are $q_{\phi_i}=-q_{\chi_j}=q$, $q_{\varphi_k}=(-1)^kq$, $q_Q=-2q$.
$X_l$ are neutral under $U(1)_X$ and vanish during inflation, their F-terms are used to stabilize $\phi_i$ and $\varphi_{1,2}$ with non-zero vacuum expectation values (VEV).

The superpotential is
\begin{equation}
\begin{split}
W=w_0+a\phi^{\frac{1}{n}}_1e^{-bT}+X_{i}(\phi_i\varphi_1-\lml)+X_0(\varphi_1\varphi_2-\lmll)  \\
+c_1(\varphi_1\varphi_4+\varphi_2\varphi_3)+m_j\varphi_4\chi_j+\varphi^2_4Q. ~~~~~~~~~~~\,  \label{sp}
\end{split}
\end{equation}
Given $q=nb\delta$, above superpotential is obviously $U(1)_X$ invariant. The first two terms in (\ref{sp}) is similar with the KKLT scenario in a gauge invariant form \cite{Kachru:2003aw}. The constant term $w_0$ is from the complex-structure moduli stabilization. Different from the KKLT scenario, the non-perturbative term $a\phi^{\frac{1}{n}}_1e^{-bT}$ is $U(1)_X$ invariant. The $U(1)_X$ transformation of matter field $\phi^{1/n}_1\rightarrow \phi^{1/n}_1e^{iq/n}$ cancels the phase factor $e^{-ib\delta}$ from shift of the modulus $T\rightarrow T+i\epsilon\delta$ under $U(1)_X$. Such kind of non-perturbative superpotential can be obtained from the gaugino condensation with massive chiral superfields, which form representation of the condensation group, such as $SU(n)$ and can be integrated out in effective field theory. The effective superpotential is guaranteed to be $U(1)_X$ invariant \cite{Taylor:1982bp,Lust:1990zi, deCarlos:1991gq, Achucarro:2006zf}. Effects of the gauge invariant non-perturbative term on moduli stabilization and inflation have been studied in \cite{Dudas:2005vv, Villadoro:2005yq, Achucarro:2006zf, Lalak:2005hr, Brax:2007fe, Lalak:2007qd}.
Furthermore, in this form the matter field $\phi_1$ has positive exponent, and gives an analytic coefficient for the non-perturbative term, which makes the anomalous $U(1)_X$ D-term cancellable. The D-term flatness
 is needed for modulus stabilization.

In Eq.~(\ref{sp}), we will have many other superpotential
terms, for example, $\phi_i \chi_j$, etc, which are allowed by
the $U(1)_X$ symmetry but neglected. To solve this problem,
 we can introduce a $Z_{m}$ discrete symmetry under which $\varphi_3$,
$\varphi_4$, $\chi_j$, and $Q$ transform as follows
\begin{equation}
\varphi_3 \rightarrow \omega^k \varphi_3~,~\varphi_4 \rightarrow \omega^{-k} \varphi_4~,~
\chi_j \rightarrow \omega^k \chi_j~,~ Q \rightarrow \omega^{2k} Q~,~\,
\end{equation}
with $\omega^m=1$, while all the other fields are neutral under $Z_m$.
Thus, superpotential will be
\begin{equation}
\begin{split}
W=w_0+a\phi^{\frac{1}{n}}_1e^{-bT}+X_{i}(\phi_i\varphi_1-\lml)+X_0(\varphi_1\varphi_2-\lmll)  \\
+c_1 \varphi_1\varphi_2+ m_0 \varphi_3\varphi_4+m_j\varphi_4\chi_j+\varphi^2_4Q~. ~~~~\,~~~~\,  \label{sp-N}
\end{split}
\end{equation}
With our numerical study assumption, we will point out  that the superpotentials in both Eq.~(\ref{sp}) and
 Eq.~(\ref{sp-N}) will give the similar inflaton potential.
In general, we can have the superpotential terms $c'_i \phi_i\varphi_1+ c_1 \varphi_1\varphi_2$.
Without loss of generality, we can make a tranformation for $\phi_i$/$\varphi_2$, and $X_i$,
and obtain the above superpotential since
$\varphi_1$ is only coupled to one linear combination of $\phi_i$ and $\varphi_2$.

\subsection*{Gauge Anomaly Cancellation}

The anomalous $U(1)_X$ plays a special role in the quantum anomaly cancellation through the Green-Schwarz mechanism
in four dimensional spacetime \cite{Dine:1987xk}. The gauge kinetic term of $U(1)_X$ is
\begin{equation}
\int d^2\theta fW^2_{\alpha}~,
\end{equation}
in which $W_\alpha$  is $U(1)_X$ gauge field strength. Here we take the gauge kinetic function $f=k_XT$. The gauge kinetic term contains two parts $Re(f)F^2$ and $Im(f)F\tilde{F}$. The first term is $U(1)_X$ invariant, while the second term transforms non-trivially under $U(1)_X$. The shift of modulus $T$ introduces an extra term $i\delta k_X\int d^2\theta W^2_{\alpha}$, which cancels the anomaly from charged fermionic fields and keep the theory anomaly free.

Ignoring the anomaly of condensation gauge group $SU(n)$, we need to consider two kinds of anomalies : the gravitational anomaly $U(1)_X$ and the cubic anomaly $U(1)^3_X$. The fermionic contributions are:
\begin{equation}
\begin{split}
{\rm Tr}~ q_z=\sum_z q_z=0 ~,~~~\,\\
{\rm Tr}~ q^3_z=\sum_z q^3_z=-6q^3.
\end{split}
\end{equation}
The gravitational anomaly is canceled without higher derivative terms $R^2$. Anomaly cancelation of cubic term $U(1)^3_X$ requires
\begin{equation}
k_X\delta=-\frac{1}{48\pi^2}\sum_z q^3_z=\frac{1}{8\pi^2}q^3.
\end{equation}
As $q=nb\delta$, we have $k_X=nbq^2/8\pi^2$.

\section{Matter Fields and Modulus Stabilization}

In this model, the matter fields are stabilized by F-terms. Even though there is the
KKLT type superpotential in (\ref{sp}), it has nothing to do with modulus stabilization but provides potential for natural inflation, actually, the real component of the modulus is stabilized by $U(1)_X$ D-term.

The F-term scalar potential is given by
\begin{equation}
V_F=e^K(K^{i\bar{j}}D_iW D_{\bar{j}}\bar{W}-3W\bar{W})~,~ \label{vks}
\end{equation}
in which $K^{i\bar{j}}$ is the inverse of the K\"ahler metric $K_{i\bar{j}}=\partial_i\partial_{\bar{j}}K$ and $D_iW=W_i+K_iW$. The complete expression of $V_F$ is rather tedious, nevertheless, it can be remarkably simplified
after field stabilization.

\subsection{Matter Fields Stabilization}

For matter fields stabilization, we ignore the constant term and the non-perturbative term, as they provide inflationary potential which is significantly lower than the matter fields stabilization scale. Their effects will be estimated later.

Clearly the matter fields $\chi_j, \varphi_{3,4}, Q$ and $X_l$ have global minimum at origin while extra matter fields $\phi_i$ and $\varphi_{1,2}$ will get non-trivial VEVs. During inflation they will evolve to the minimum rapidly driven by the exponential factor $e^K$ of F-term scalar potential and the large masses obtained from the matter couplings in
 Eq.~(\ref{sp}).
Therefore the value of superpotential during inflation is simplified as $\langle W\rangle \equiv W_0= w_0+a\phi^{\frac{1}{n}}_1e^{-bT}$. In Eq. (\ref{vks}), only these terms independent with $\chi_j, \varphi_{3,4}, Q$ and $X_l$ are non-vanishing. The non-vanishing F-terms $V_z$ introduced by fields $z$ are (multiplied by an overall factor $e^K$):

For $T$:
\begin{equation}
\begin{split}
V_T=(T+\bar{T})^2[a^2b^2(\phi_1\bar{\phi_1})^{\frac{1}{n}}e^{-b(T+\bar{T})}
+2a^2b\frac{1}{T+\bar{T}}(\phi_1\bar{\phi_1})^{\frac{1}{n}}e^{-b(T+\bar{T})} \\
+abw_0\frac{1}{T+\bar{T}}(\phi^{\frac{1}{n}}e^{-bT}+\bar{\phi}^{\frac{1}{n}}e^{-b\bar{T}})]+W_0\bar{W}_0, \label{vf1}
\end{split}
\end{equation}
for $\phi_1$:
\begin{equation}
\begin{split}
V_{\phi_1}=\frac{a^2}{n^2}(\phi_1\bar{\phi_1})^{\frac{1}{n}-1}e^{-b(T+\bar{T})}
+\frac{2a^2}{n}(\phi_1\bar{\phi_1})^{\frac{1}{n}}e^{-b(T+\bar{T})} \\ + \frac{a}{n}w_0 (\phi^{\frac{1}{n}}e^{-bT}+\bar{\phi}^{\frac{1}{n}}e^{-b\bar{T}})+\phi_1\bar{\phi_1}W_0\bar{W}_0,\label{vf2}
\end{split}
\end{equation}
for $\phi_{i>1}$:
\begin{equation}
V_{\phi_{i>1}}=\phi_{i>1}\bar{\phi}_{i>1}W_0\bar{W}_0,\label{vf3}
\end{equation}
and similarly, for $\varphi_1$ and $\varphi_2$:
\begin{equation}
V_{\varphi_{1,2}}=(\varphi \bar{\varphi})_{1,2}W_0\bar{W}_0.\label{vf4}
\end{equation}
Above terms are proportional to $a^2$, $aw_0$ or $w^2_0$. In this model, the parameters have orders of $a\sim w_0\sim10^{-5}$, while $c_1\sim 10^{-3}-10^{-2}$, therefore above terms are extremely small comparing with the following F-terms.

For the superpotential in Eq.~(\ref{sp-N}), $V_T$, $V_{\phi_1}$, $V_{\phi_{i>1}}$ are the same, while $V_{\varphi_{1,2}}$
become
\begin{equation}
V_{\varphi_{1,2}}= |\bar{\varphi} W_0 + c_1 \varphi_{2,1}|^2~.~
\end{equation}
Note that in the above equation
only the terms $|c_1 \varphi_{1,2}|^2$ are relevant while all the other terms are small and  negligible,
thus, all the rest discussions will be the same and we will not repeat here.

The vanishing neutral matter fields $X_l$ provide F-terms $|W_{X_l}|^2$ in the scalar potential, besides, $\varphi_1$ and $\varphi_2$ get mass terms from $|W_{\varphi_4}|^2$ and $|W_{\varphi_3}|^2$, respectively. Combining these F-terms together we have
\begin{equation}
V_{F_1}=e^K|W_{X_l}|^2=e^K (\sum_i|\phi_i\varphi_1-\lml|^2+|\varphi_1\varphi_2-\lmll|^2+c^2_1(|\varphi_1|^2+|\varphi_2|^2)).\label{vf}
\end{equation}
Of course there are numerous corrections containing $\phi_i$ and $\varphi_{1,2}$ in (\ref{vf1}, \ref{vf2}, \ref{vf3}, \ref{vf4}), however, they are either significantly smaller than (\ref{vf}) or can be canceled with each other. The VEVs of $\phi_i$ are mainly dominated by the results obtained from (\ref{vf}).

Taking $\varphi_i=r_ie^{i\theta_i}$, the potential (\ref{vf}) becomes
\begin{equation}
\begin{split}
V_{F_1}=e^K|W_{X_l}|^2=e^K \{\sum_i|\phi_i\varphi_1-\lml|^2+r^2_1r^2_2-2\lmll r_1r_2{\rm cos}(\theta_1+\theta_2)+\lmll^2+
c^2_1(r^2_1+r^2_2)\} \\
\geqslant e^K \{\sum_i|\phi_i\varphi_1-\lml|^2+r^2_1r^2_2-2\lmll r_1r_2+\lmll^2+c^2_1(r^2_1+r^2_2)\} ~~~~~~~\;~~~~~~~~~~~~~~~~~\\
=e^K \{\sum_i|\phi_i\varphi_1-\lml|^2+(r_1r_2-(\lmll-c^2_1))^2+c^2_1(r_1-r_2)^2+c^2_1(2\lmll-c^2_1)\}. ~~~~~
\end{split} \label{vs}
\end{equation}
The vacuum takes place at $\theta_1+\theta_2=2n\pi$, while the direction $\theta_1-\theta_2$ is flat.
The minimization of the potential in the bracket gives
\begin{equation}
\begin{split}
r_1=r_2=r_0=\sqrt{\lmll-c^2_1}, \\
|\phi_i|\equiv r=\frac{\lml}{r_0}.~~~~~~~~~~~~~~~~
\end{split} \label{vev}
\end{equation}
The orders of parameters are simply taken as $r_0\sim10^{-2}$ and $r^2=0.1$ in following estimations. Eq. (\ref{vs}) provides strong stabilization on the matter fields. For the condition $r_1=r_2$, the mass of direction along $(r_1-r_2)/\sqrt{2}$ is $2c_1\sim 10^{-3}-10^{-2}\gg H$, where $H$ is Hubble constant during inflation, therefore it is strongly stabilized as well.
Besides, we get a cosmology constant term $V_0=c^2_1(2\lmll-c^2_1)$.

The anomalous $U(1)_X$ is spontaneously broken by VEVs of charged matter fields. The phases of $\phi_i=re^{i\alpha_i}$ satisfy $\alpha_i=\theta_2$, we take the $U(1)_X$ gauge $\theta_2=0$ for simplicity.
The flat direction is absorbed by $U(1)_X$ massive vector field through the Higgs mechanism after spontaneous symmetry broken.

It should be careful to consider above values as VEVs of these fields. Actually the cosmology constant term $V_0$, together with the overall factor $e^K$, can shift the non-zero VEVs. Specifically, the first condition $r_1=r_2$ remains the same, while the results of $\phi_i$ and $r_0=\sqrt{\lmll-c^2_1}$ will be slightly modified.

In general, considering a field $\phi$ with mass $m_0$ and non-zero VEV $\phi_0$, assuming its K\"ahler potential is minimal, and the overall scalar potential $V=e^{\phi^2}(\frac{1}{2}m^2_0(\phi-\phi_0)^2+V_0)$, where $V_0$ is the residual
cosmology constant term after field stabilization or during inflation, the vacuum is determined by
\begin{equation}
\frac{dV}{d\phi}=e^{\phi^2}(m^2_0(\phi-\phi_0)+ m^2_0 \phi (\phi-\phi_0)^2+2V_0\phi)=0,
\end{equation}
which gives a new VEV $\phi'_0\simeq \phi_0-2V_0\phi_0/m^2_0$. During inflation, the quasi-cosmology constant provided by inflaton is $V_0\sim 10^{-8}$ in the Planck units, $m_0$ is of order $O(10^{-2})$ in our model, therefore the shift of matter fields is about $\Delta\phi/\phi_0\simeq10^{-4}$. The vacuum energy is reduced by $2V^2_0\phi^2_0/m^2_0\sim10^{-13}$, which is completely ignorable during inflation. Therefore we can safely consider the values obtained in (\ref{vev}) are VEVs of these fields. Besides, there is a coupling between $\phi_1$ and modulus $T$ through $a\phi^{\frac{1}{n}}_1e^{-bT}$, however, we will show that the interaction is seriously suppressed by $aw_0\sim10^{-9}$ and has ignorable effect on $\phi_1$ stabilization as well.

\subsection{F-term Potential after Stabilization}

After field stabilization, the F-term potential is simplified. Besides an overall factor $e^K$, it is
\begin{equation}
\begin{split}
V=(T+\bar{T})^2[a^2b^2(\phi_1\bar{\phi_1})^{\frac{1}{n}}e^{-b(T+\bar{T})}
+2a^2b\frac{1}{T+\bar{T}}(\phi_1\bar{\phi_1})^{\frac{1}{n}}e^{-b(T+\bar{T})} ~\\
+abw_0\frac{1}{T+\bar{T}}(\phi^{\frac{1}{n}}_1e^{-bT}+\bar{\phi}^{\frac{1}{n}}_1e^{-b\bar{T}})]
+\frac{a^2}{n^2}(\phi_1\bar{\phi_1})^{\frac{1}{n}-1}e^{-b(T+\bar{T})} \,\\
+\frac{2a^2}{n}(\phi_1\bar{\phi_1})^{\frac{1}{n}}e^{-b(T+\bar{T})} + \frac{a}{n}w_0 (\phi^{\frac{1}{n}}_1e^{-bT}+\bar{\phi}^{\frac{1}{n}}_1e^{-b\bar{T}}) ~~~~~~~~~~~~\\
+c^2_1(2\lambda_2-c^2_1)+(\sum_i|\phi_i|^2+|\varphi_1|^2+|\varphi_2|^2-2)W_0\bar{W}_0. ~~~~~~\label{v0}
\end{split}
\end{equation}
Taking $r^2_0\ll r^2$, we ignore the contributions from $|\varphi_{1,\, 2}|^2W_0\bar{W}_0$. The parameters can be simply taking as $\sum_i|\phi_i|^2=mr^2=2$ so that the last term in Eq. (\ref{v0}) vanishes.

Couplings between modulus $T$ and $\phi_1$ are shown in Eq. (\ref{v0}), the dominant term is $g(T)\phi^{\frac{1}{n}}_1$ with $g(T)\sim 10^{-8}$. Ignoring the factor $e^K$, the potential of $\phi_1$ is approximate to
\begin{equation}
V_{\phi_1}=\frac{1}{2}m^2_1(\phi_1-r)^2+g(T)\phi^{\frac{1}{n}}_1,
\end{equation}
in which $m^2_1\sim10^{-4}$. VEV of $\phi_1$ is shifted about $\Delta\langle\phi_1\rangle\simeq g(T)r^\frac{1}{n}/nm^2_1r\sim10^{-4}$, and the energy is reduced by $\frac{1}{2}\frac{g(T)^2r^{2/n}}{n^2m^2_1r^2}\sim10^{-12}$, which confirms that the non-perturbative effect is ignorable for matter field stabilization.

\subsection{Anomalous $U(1)_X$ D-term and Modulus Stabilization}

The anomalous $U(1)_X$ D-term scalar potential is given by
\begin{equation}
V_D=\frac{1}{2{\rm Re}(f)}D^2,
\end{equation}
 where $D=iK_zX^z+i\frac{W_z}{W}X^z$. $X^z$ are the components of Killing vector corresponding to $U(1)_X$ isometries of the K\"ahler manifold, which are
\begin{equation}
(X^T,\; X^{\phi_i},\; X^{\chi_j}, \;X^{\varphi_k}, \; X^Q)=(i\delta,\; iq\phi_i ,\; -iq\chi_j, \;(-1)^kiq\varphi_k,\;-2iqQ).
\end{equation}
In this model the superpotential is gauge invariant, the D-term is simplified as $D=iK_zX^z$
and it reads
\begin{equation}
D=\frac{\delta}{T+\tb}-q\sum_i|\phi_i|^2+q\sum_j|\chi_j|^2-(-1)^kq\sum_k|\varphi_k|^2+2q|Q|^2.
\end{equation}
After field stabilization, $\chi_j, \varphi_{3,4}$ and $Q$ have vanshing VEVs, the $\varphi_1$
and $\varphi_2$ D-terms cancel each other due to
$|\varphi_1|=|\varphi_2|$, and then the D-term is reduced to
\begin{equation}
D=\frac{\delta}{T+\tb}-q\sum_i|\phi_i|^2=\frac{\delta}{T+\tb}-mqr^2~.
\end{equation}
The D-term potential is
\begin{equation}
\begin{split}
V_D=\frac{1}{2k_XT_R}(\frac{\delta}{T+\tb}-mqr^2)^2 ~\\
=\frac{4\pi^2}{(nb)^3T_R}(\frac{1}{2T_R}-mnbr^2)^2,
\end{split}
\end{equation}
in which $T_R\equiv{\rm Re}\,T$ and the gauge invariant condition $q=nb\delta$ and cubic anomaly $U(1)^3_X$ cancelation $k_X=nbq^2/8\pi^2$ are used.
The  D-term vanishing condition gives $\langle T_R\rangle\equiv T_0=1/2mnbr^2$. In the simplified case
with $mr^2=2$, the real component of modulus is $T_0=1/4nb$.

\section{Natural Inflation Potential}

We have stabilized all the fields except the imaginary component of $T$. Now the scalar potential (\ref{v0}) becomes
\begin{equation}
\begin{split}
V=2e^2b\,[a^2 r^{\frac{2}{n}}e^{-\frac{1}{2n}}(\frac{1}{4n}+3+\frac{1}{nr^2}) ~~~~~~~~\\
+nc^2_1(2\lambda_2-c^2_1)+ 3aw_0r^{\frac{1}{n}}e^{-\frac{1}{4n}}{\rm cos}(b\theta)],
\end{split}
\end{equation}
where $\theta$ is the imaginary component of $T$.
To get global Minkowski vacuum the parameters need to be adjusted so that
\begin{equation}
nc^2_1(2\lambda_2-c^2_1)+a^2 r^{\frac{2}{n}}e^{-\frac{1}{2n}}(\frac{1}{4n}+3+\frac{1}{nr^2})=3aw_0r^{\frac{1}{n}}e^{-\frac{1}{4n}}, \label{pa}
\end{equation}
and we get the scalar potential
\begin{equation}
V=6abw_0e^2r^{\frac{1}{n}}e^{-\frac{1}{4n}}\,(1+{\rm cos}(b\theta)),
\end{equation}
which is of the same form in Eq. (\ref{V}) with $\Lambda^4=6abw_0e^2r^{1/n}e^{-1/4n}$. Taking $a\simeq 5\times10^{-5}$, $w_0\sim6\times 10^{-5}$, $r^2\sim0.1$, $b=0.1$ and $n\geqslant6$, it gives the inflation scale $\Lambda\sim10^{-2}$ in the Planck units.

Before we consider $\theta$ as our inflaton, a field re-scale is needed to get canonical kinetic term, and this will affect the decay constant of $\theta$.

The kinetic term of $T$ is
\begin{equation}
L_K=\frac{1}{(T+\bar{T})^2}\partial_{\mu}T\partial^{\mu}T=\frac{1}{4T^2_0}(\partial_{\mu}T_R\partial^{\mu}T_R
+\partial_{\mu}\theta\partial^{\mu}\theta).
\end{equation}
Defining $\theta=\sqrt{2}T_0\rho$, we get the action of inflaton
\begin{equation}
L=\frac{1}{2}\partial_{\mu}\rho\partial^{\mu}\rho+\Lambda^4(1+{\rm cos}(\frac{\rho}{\sqrt{2}mnr^2})),
\end{equation}
in which the D-term stabilization condition $bT_0=1/2mnr^2$ has been used. The axion decay constant in this model is $f=\sqrt{2}mnr^2$. The VEV of matter fields $r$ is smaller than the Planck mass. Without uplift from parameters $m, n$, the axion decay constant $f$ cannot be super-Planckian. Nevertheless, $f$ is proportional to the product of charged field number $m$ and the degree of condensation gauge group $n$, it is very easy to get super-Plankian $f$ by taking large $m$ or $n$. We used $r^2=0.1$ and $mr^2=2$ before, in such case we have $m=20$, then $f=2\sqrt{2}n$ and it is of order $O(10)$ with $n=4$. By using larger $n$ the axion decay constant $f$ increases linearly, the potential gets close to the type $\frac{1}{2}m^2\rho^2$, and we get chaotic inflation.

\subsection*{Gravitino Mass}

Even though the matter fields are stabilized at the scales much higher than inflation scale,
they do not introduce too heavy gravitino mass.   After field stabilization, the pure matter couplings
in the superpotential in Eq.~(\ref{sp}) vanish, the VEV of superpotential $\langle W\rangle$
during inflation is
\begin{equation}
\langle W\rangle\equiv W_0=w_0+a\langle\phi^{\frac{1}{n}}_1e^{-bT}\rangle=w_0-ar^{\frac{1}{n}}e^{-\frac{1}{4n}}.
\end{equation}
Besides, we also have the VEV of $e^K$
\begin{equation}
\langle e^K\rangle=\frac{e^2}{2T_0},
\end{equation}
in which we have used $mr^2=2$ and the small term $2r^2_1$ from VEVs of $\varphi_{1,2}$ is ignored. The gravitino mass is
\begin{equation}
\begin{split}
M_{\frac{3}{2}}\equiv{\rm exp}(\frac{\langle G\rangle}{2})=\langle e^{\frac{K}{2}}(W_0\bar{W}_0)^{\frac{1}{2}}\rangle ~~~~~~~~~~~\,\\
=e\sqrt{2nb}(w_0-ar^{\frac{1}{n}}e^{-\frac{1}{4n}}).
\end{split}
\end{equation}
The gravitino mass relates to the inflation energy scale through
\begin{equation}
(e\sqrt{2nb})^2w_0ar^{\frac{1}{n}}e^{-\frac{1}{4n}}=\frac{n}{3}\Lambda^4.
\end{equation}
Small gravitino mass can be obtained by taking $w_0\rightarrow ar^{\frac{1}{n}}e^{-\frac{1}{4n}}$, however, Eq. (\ref{pa}) provides a lower bound on it
\begin{equation}
\begin{split}
w_0-ar^{\frac{1}{n}}e^{-\frac{1}{4n}}=nc^2_1(2\lmll-c^2_1)(3ar^{\frac{1}{n}}e^{-\frac{1}{4n}})^{-1}
+\frac{1}{3n}ar^{\frac{1}{n}}e^{-\frac{1}{4n}}(\frac{1}{4}+\frac{1}{r^2})   \\
\geqslant \frac{2}{3}c_1(2\lmll-c^2_1)^{\frac{1}{2}}(\frac{1}{4}+\frac{1}{r^2})^{\frac{1}{2}}. ~~~~~~~~~~~~~~~~~~~~~~~~~~~~~~~
\end{split}
\end{equation}
The minimum locates at $c^2_1(2\lmll-c^2_1)\simeq (\frac{1}{4}+\frac{1}{r^2})w^2_0/n^2$. For $c_1\sim 10^{-3}$, $\lmll\sim10^{-4}$, the gravitino mass can be reduced to the order of $10^{-5}$, while it cannot get significantly smaller otherwise the field stabilization is not strong enough for inflation. In short,
the gravitino mass will not affect inflation in our model.

\section{Conclusion}

We have proposed a natural infaltion model based on string inspired SUGRA with gauged shift symmetry
$U(1)_X$.
The matter fields are stabilized by F-terms, part of them obtain non-zero VEVs which break the anomalous $U(1)_X$ spontaneously. They obtain masses several orders larger than the mass of inflaton, consequently the effect of vacuum energy from inflaton on field stabilization is seriously suppressed and ignorable. The
modulus-dependent FI term of $U(1)_X$ plays a critical role in modulus stabilization. As the coupling between matter field and modulus is analytic, the D-term can be vanished in our model. Once the matter fields obtain non-zero VEVs, the real component of modulus $T$ is fixed by the D-term flatness. While the D-term is independent with the axionic component of $T$, therefore its cancellation has no effect on axion which remains light after modulus stabilization. The anomalous $U(1)_X$ splits the masses of real and imaginary components of $T$. Such role of anomalous $U(1)_X$ D-term on modulus stabilization has been studied in \cite{Li:2014owa}. In the F-term moduli stabilization, usually the imaginary component of modulus obtains mass as large as the real component \cite{Kallosh:2007ig}, the axion inflation cannot be realized. The reason is in F-term potential, the real and imaginary components of moduli couple with each other, it is highly non-trivial to split their masses at different scale so that one of them can play the role of inflaton while the other
 is frozen during inflation.

Potential for natural inflation is obtained from non-perturbative effect. Generally it is very difficult to get trans-Planckian axion decay constant, which is needed to fit with BICEP2 results. In our model, the axion decay constant linearly depends on the degree of condensation gauge group $SU(n)$ and number of $U(1)_X$ charged matter fields, therefore, the super-Planckian axion decay constant can be easily fulfilled by using large condensation gauge group and more $U(1)_X$ charged fields.

\begin{acknowledgments}

The work of DVN was supported
in part by the DOE grant DE-FG03-95-ER-40917. The work of TL is supported in part by
by the Natural Science
Foundation of China under grant numbers 10821504, 11075194, 11135003, and 11275246, and by the National
Basic Research Program of China (973 Program) under grant number 2010CB833000.

\end{acknowledgments}

\end{document}